%% file: CIKM21.tex
\documentclass[sigconf]{acmart}
\setcopyright{acmcopyright}
\acmConference[CIKM '21]{CIKM '21: ACM International Conference on Information and Knowledge Management}{November 01--05, 2021}{Virtual Event, Australia}

\copyrightyear{2021}
\acmYear{2021}
\setcopyright{acmcopyright}\acmConference[CIKM '21]{Proceedings of the 30th ACM
International Conference on Information and Knowledge Management}{November
1--5, 2021}{Virtual Event, QLD, Australia}
\acmBooktitle{Proceedings of the 30th ACM International Conference on Information
and Knowledge Management (CIKM '21), November 1--5, 2021, Virtual Event, QLD,
Australia}
\acmPrice{15.00}
\acmDOI{10.1145/3459637.3482351}
\acmISBN{978-1-4503-8446-9/21/11}

\usepackage{xcolor}
\usepackage{bm}
\usepackage{balance}
\usepackage[ruled,linesnumbered]{algorithm2e}
\usepackage{amsfonts}
\usepackage{amsmath}
\usepackage{graphicx}
\usepackage{subcaption}
\usepackage{threeparttable}
\usepackage{booktabs}
\usepackage{multirow}
\usepackage{caption}
\usepackage{xr}
\newtheorem{problem}{Problem}

\begin{document}

\title{Hyperbolic Hypergraphs for Sequential Recommendation}










\author{Yicong Li}
\affiliation{
 \institution{University of Technology Sydney}
 \country{Australia}}
\email{Yicong.Li@student.uts.edu.au}

\author{Hongxu Chen}
\authornote{Having equal contribution with the first author.}
\authornote{Corresponsing authors.}
\affiliation{
  \institution{University of Technology Sydney}
  \country{Australia}
}
\email{Hongxu.Chen@uts.edu.au}

\author{Xiangguo Sun}
\affiliation{
  \institution{Southeast University}
  \country{China}
}
\email{sunxiangguo@seu.edu.cn}

\author{Zhenchao Sun}
\affiliation{
  \institution{Shandong University}
  \country{China}
}
\email{zhenchao.sun@mail.sdu.edu.cn}

\author{Lin Li}
\affiliation{
  \institution{Wuhan University of Technology}
  \country{China}
}
\email{cathylilin@whut.edu.cn}

\author{Lizhen Cui}
\affiliation{
  \institution{Shandong University}
  \country{China}
}
\email{clz@sdu.edu.cn}

\author{Philip S. Yu}
\affiliation{
 \institution{University of Illinois at Chicago}
 \country{the United States}}
 \email{psyu@cs.uic.edu}

\author{Guandong Xu}\authornotemark[2]
\affiliation{
 \institution{University of Technology Sydney}
 \country{Australia}}
\email{Guandong.Xu@uts.edu.au}

\begin{abstract}

Hypergraphs have been becoming a popular choice to model complex, non-pairwise, and higher-order interactions for recommender system. However, compared with traditional graph-based methods, the constructed hypergraphs are usually much sparser, which leads to a dilemma when balancing the benefits of hypergraphs and the modelling difficulty. Moreover, existing sequential hypergraph recommendation overlooks the temporal modelling among user relationships, which neglects rich social signal from the recommendation data. To tackle the above shortcomings of the existing hypergraph-based sequential recommendations, we propose a novel architecture named \textbf{H}yperbolic \textbf{H}ypergraph representation learning method for \textbf{Seq}uential \textbf{Rec}ommendation ($\mathrm{H^2SeqRec}$) with pre-training phase. Specifically, we design three self-supervised tasks to obtain the pre-training item embeddings to feed or fuse into the following recommendation architecture (with two ways to use the pre-trained embeddings). In the recommendation phase, we learn multi-scale item embeddings via a hierarchical structure to capture multiple time-span information. To alleviate the negative impact of sparse hypergraphs, we utilize a hyperbolic space-based hypergraph convolutional neural network to learn the dynamic item embeddings. Also, we design an item enhancement module to capture dynamic social information at each timestamp to improve effectiveness. Extensive experiments are conducted on two real-world datasets to prove the effectiveness and high performance of the model.


\end{abstract}

\maketitle
\pagestyle{plain}
\input{intro}
\input{problem-def}
\input{model}

\input{experiment}
\input{related-work}

\vspace{-1em}
\section{Conclusion}\label{sec:con}
In this work, we focus on the sparse problem of most existing hypergraph-based sequential recommendation and on the lacking exploitation of hidden hyperedges among users problem, and propose \textbf{H}yperbolic \textbf{H}ypergraph representation learning method for \textbf{Seq}uential \textbf{Rec}ommendation ($\mathrm{H^2SeqRec}$) with pre-training phase. Experiments show that our proposed model outperforms the state-of-the-art sequential recommendations and each of the components contributes to the whole architecture. However, on a sparse dataset, the data augmentation improves the recommendation, while on a dense dataset, it is useless and even worse, which proves dense datasets do not need data augmentation to achieve better performance. In future work, we will explore multiple sequential ways to model the sequential recommendation data on graphs.

\section*{Acknowledgment}
This work is supported by the Australian Research Council (ARC)
under Grant No. DP200101374 and LP170100891, and NSF under grants III-1763325, III-1909323,  III-2106758, and SaTC-1930941.

\bibliographystyle{abbrv}
\balance
\bibliography{ref.bib}
\end{document}

%% file: intro.tex
\vspace{-1em}
\section{Introduction}
\label{sec:introudction}
\begin{figure}[htp]
	\centering
	\includegraphics[width=0.45\textwidth]{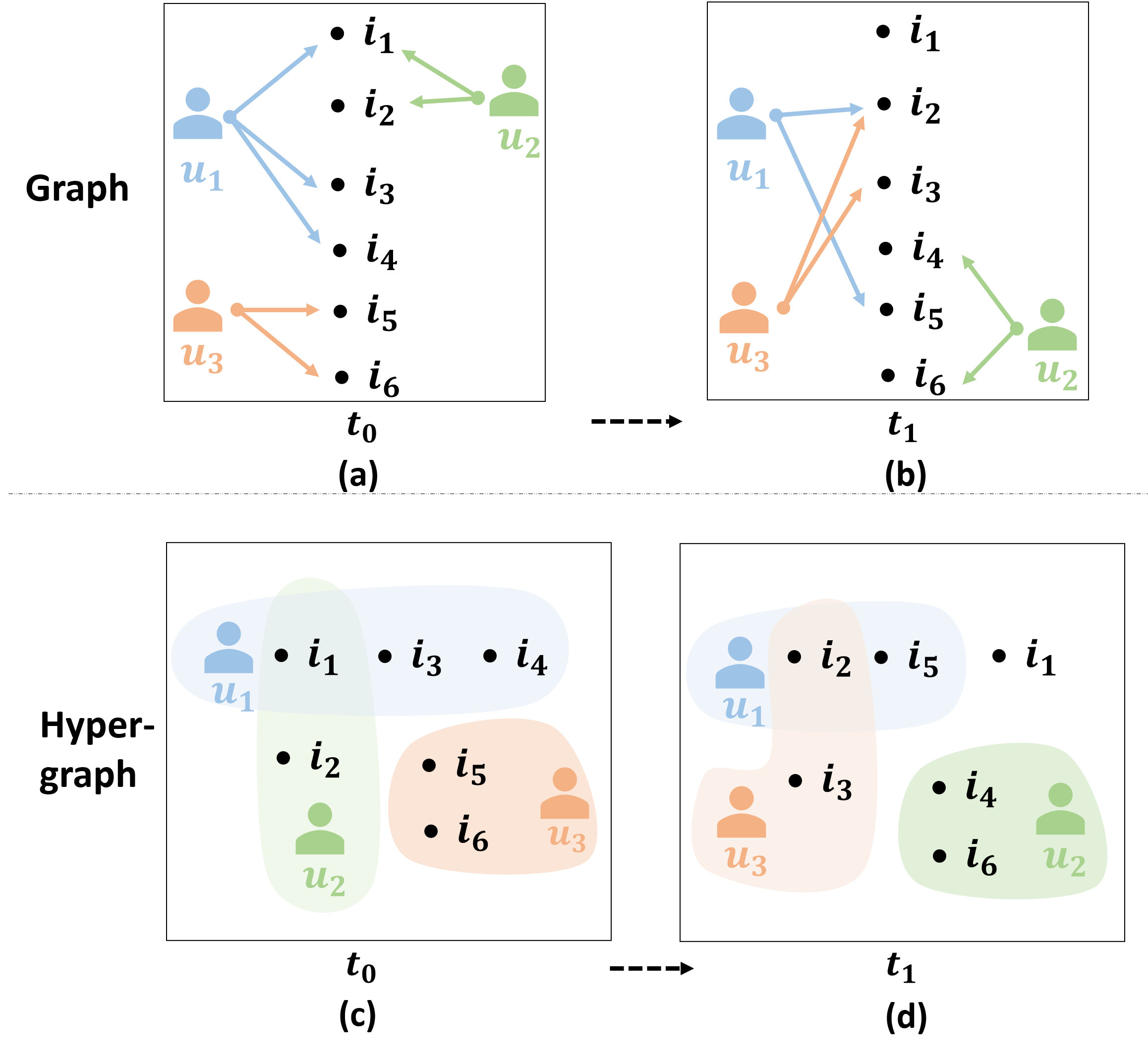}
	\vspace{-2em}
	\caption{A comparison of sequential graph construction and hypergraph construction. Single-colored area in (c) and (d) denote hyperedges.} \vspace{-3em}
	\label{fig:toy_example}

\end{figure}

Graph-based approaches have been widely used and achieved great improvement for next-item recommender systems. However, most existing literature \cite{yuan2014graph, wang2020personalized, kang2018self, sun2019bert4rec, wu2019session, DBLP:conf/kdd/MaKL19, wang2020next, qin2020sequential, chen2021temporal} treat the dynamic time-dependent user-item interactions as a temporal bipartite graph and learn their latent representations for action. Though the graph-based graph modelling can capture the beneficial first-order (i.e., user-item interactions) and second-order (i.e., co-purchasing) interactions for recommendation, the higher-order signals among users and items are usually neglected by existing works due to the limitations of traditional graph modelling. With noting the shortcoming, recent works \cite{wang2020next, xia2020self, wang2021session, gharahighehi2020fair, yu2021self, sun2021multi, sun2021heterogeneous} resorted hypergraphs to make up and developed hypergraph-based modelling approaches for sequential recommender systems. The basic idea is illustrated in Figure \ref{fig:toy_example}, when using traditional graphs to model user-item dynamic relationships evolution, the learned information from Figure \ref{fig:toy_example}.(a) to \ref{fig:toy_example}.(b) is monotonous due to the simple pair-wise data structure. In contrast, hypergraphs are able to capture high-order dynamic relationships (buy items at the same time, similar user groups, etc.) thanks to the non-pairwise data structure. That is, unlike traditional graphs, an edge in the hypergraph (a.k.a hyperedge) is allowed to connect more than two nodes, promising to capture multi-scale signals for the recommendation.

Nevertheless, all these existing works ignored a critical issue of such hypergraph-based approaches, which is \textbf{the sparsity issue is becoming more severe in hypergraph based recommender systems}.
As analysed in \cite{yin2020learning}, the recommendation benchmark Amazon dataset is sparse and exhibits long-tailed distribution, in which many users have limited interactions to the items, while only a small number of users interact with many items. In such a case, the constructed hypergraphs from the original dataset will be much sparser, resulting in insufficient training samples for action. To be specific, if we construct hypergraphs based on the original simple user-item bipartite graph, the number of items (nodes) does not change, but the number of hyperedges is dramatically shrunken compared to the number of original links between users and items, which is obvious in Figure \ref{fig:toy_example}. Therefore, when constructing hypergraphs, the challenging sparsity issue and long-tailed data distribution are becoming even more severe. 

The second limitation of existing hypergraph-based sequential recommendation lies in \textbf{lacking exploitation of hidden hyperedges among users}, which we believe will be beneficial to understand the hidden but insightful behaviours among users (e.g., common interests groups, co-purchasing groups, users who have similar buying patterns, etc.). Taking the group recommendation task \cite{wang2020group, yin2019social} as an example, it focuses on a group of users' preference, which means users in the same group may tend to have a similar preference, or at least they have more common interest compared to the rest of the world. Inspired from the idea, we are curious about the possibility of exploring and leveraging hypergraphs constructed from users side to improve the overall recommendation performance. For example, in Figure \ref{fig:toy_example}(c) and (d), from the timestamp $t_0$ to $t_1$, they are the evolution of sequential hypergraph constructions. It is obvious that in the timestamp $t_0$, the user $u_1$ is of co-purchasing relationship with $u_2$, and the hyperedges of $u_1$ and $u_2$ are connected. In this case, the items $i_1$, $i_2$, $i_3$ and $i_4$ are likely to be more similar to each other than to $i_5$ and $i_6$. The same is also in the timestamp $t_1$. Therefore, the aim is to capture the dynamic items diversification through dynamic hypergraphs constructions according to users' relationships to enhance the model.
Intuitively, like traditional graph-based methods, this try may gain improvement on recommendation. 



To holistically solve the above issues, we propose a novel architecture named \textbf{H}yperbolic \textbf{H}ypergraph representation learning for \textbf{Seq}uential  \textbf{Rec}ommendation ($\mathbf{H^2SeqRec}$).
Specifically, we propose to firstly pre-train the model with self-supervised learning on three well-sophisticated tasks. It is worth noting that compared with the prior works paying more attention to chronologically model each user's buying history, one of our contributions is to introduce a hyperedge prediction task, which explicitly models the users' historical records as hyperedges at each timestamp and explores the potential relationships between linked hyperedges (users who have similar buying history) \cite{zhou2020s3, xie2020contrastive}. Besides, we also investigate hyperbolic embedding spaces \cite{zhang2021we} and manage to map the sparse data points to the hyperboloid manifold directly. The rationale is that hyperbolic space has a stronger ability than Euclidean space to accommodate networks with long-tailed distributions and sparse structures \cite{ganea2018hyperbolic, chami2019hyperbolic, li2021hyperbolic, song2021hyperbolic, luo2020interpretable}, which is also verified in our experiments.




In addition, to exploit hidden but useful user-side information in sequential recommendation settings, we propose to construct an induced hypergraph for each user to model his behaviour pattern in group-level. Intuitively, a user's behaviour pattern is not only reflected in his historical records but can also be excavated from other users who have similar behaviour pattern. This is particularly helpful to remit the "cold start" problem. Specifically, for the target user $u$, we first use a hyperedge to collect historical items taken by $u$ at the timestamp $t$. Then we find the other hyperedges from other users who share the overlapping items with $u$ at the timestamp $t$. All these hyperedges construct an induced hypergraph to describe the user $u$'s behaviour pattern from the group view at the timestamp $t$. 


\begin{figure}[t]
	\centering
	\includegraphics[width=0.45\textwidth]{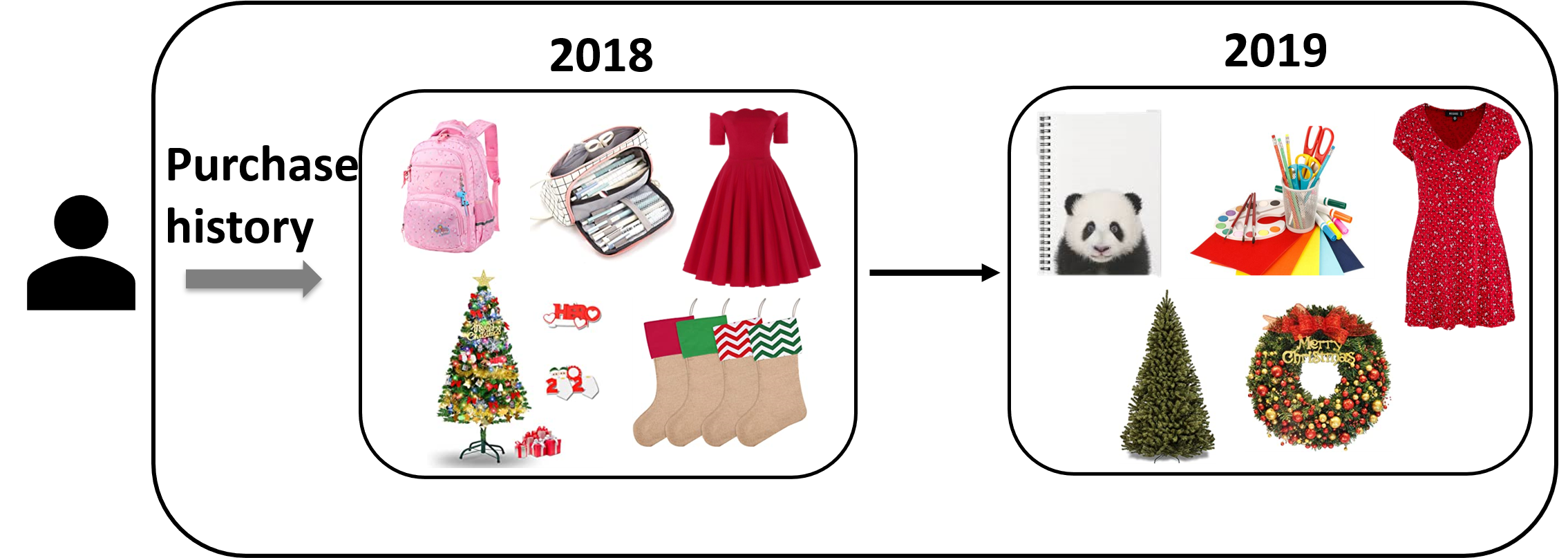}\vspace{-1em}
	\caption{An example of yearly purchase behaviour.} \vspace{-2.5em}
	\label{fig:toy_example2}
\end{figure} 

Although constructing hypergraphs from original simple user-item bipartite graphs seems a better choice, it is difficult to decide the time granularity to which the length of item sequence for each hyperedge is split. After analysing the items in the dataset, we find the purchase history is not only in a chronological sequence way, but with a periodic regularity, like seasonal evolution. As shown in Figure \ref{fig:toy_example2}, the user bought a Christmas tree and some decorations in both 2018 and 2019, and she also bought two dresses in the summer of 2018 and 2019. Therefore, rather than model each user's items in a chronological way, we design three views for hypergraph construction, namely yearly, quarterly and monthly views, and manage to hierarchically learn latent item representations. Moreover, we also find that when constructing quarterly and yearly hypergraphs, the hypergraphs are no longer as sparse as original single-time grained hypergraphs. In this way, the hierarchical time-span method for hypergraph construction also alleviates the above mentioned sparse problem.

In summary, the contributions of this work are as follows.


\begin{itemize}
	
	
	
	\item We propose three self-supervised learning tasks as the pre-training phase. To our best knowledge, we are the first to propose a hypergraph link prediction as a pre-training task to do data augmentation.
	
	\item We explore hidden insightful behaviours among users by constructing hypergraphs related to users at each timestamp to enhance the recommendation model.
	
	\item Instead of traditional chronological sequential modelling of data, we argue periodical regularity and model users' interest as a hierarchical structure for improving recommendation.



	\item We propose a novel \textbf{H}yperbolic \textbf{H}ypergraph representation learning method for \textbf{Seq}uential \textbf{Rec}ommendation \\($\mathbf{H^2SeqRec}$) to well-model long-tail data in sequential recommendation task. 
	
	\item Extensive experiments are conducted on benchmark datasets. Our model outperforms the SOTA sequential recommendations, which shows the effectiveness of our model. 

\end{itemize}



%% file: problem-def.tex
\externaldocument{intro}
\vspace{-1em}
\section{Problem Formulation}
The section presents the definition and the problem formulation.

\newtheorem{definition3}{Definition}

\begin{definition3}
	\textbf{Hypergraph.} Assuming $\mathcal{G}^H=(\mathcal{V},\mathcal{E}^H)$ denotes a hypergraph with a nodes set $\mathcal{V}$ and a hyperedges set $\mathcal{E}^H$, a hyperedge $e^H\in \mathcal{E}^H$ connects multiple $n$ nodes ($n\geq 2$). The nodes set connected by a hyperedge $e^H$ is a hypernode $v^H \subset \mathcal{V}$.
\end{definition3}

\vspace{-1em}
\begin{problem}
	\textbf{Sequential Recommendation.} In the sequential recommendation task, the user's related items are in chronological order. Given the user set $\mathcal{U} = \{u_1, u_2, ..., u_n\}$ and the item set $\mathcal{I} = \{i_1, i_2, ..., i_m\}$, each user has sequential items $\mathcal{S}_{u_n} = \{s_{u_n}^1, s_{u_n}^2, ..., s_{u_n}^t\}$, where $s_{u_n}^t \in \mathcal{I}$ denotes the $t^{th}$ historical item of user $u_n$. The sequential recommendation problem is to predict the next $q$ items $\{s_{u_n}^{t+1}, ..., s_{u_n}^{t+q}\}$ associated to each user $u$, according to sequential historical items $\mathcal{S}_{\mathcal{U}} = \{\mathcal{S}_{u_1}, \mathcal{S}_{u_2}, ..., \mathcal{S}_{u_n}\}$. In this paper, $q=1$. For each timestamp $t$, a hypergraph $\mathcal{G}^H_t$ is constructed according to the current user set $\mathcal{U}^t$ and the item set $\mathcal{I}^t$. Each user's historical items within the timestamp $t$ are connected by a hyperedge.
\end{problem}
\vspace{-3em}

%% file: model.tex
\section{Model}

\label{sec:model}


\begin{figure}[t]
	\centering
	\includegraphics[width=0.45\textwidth]{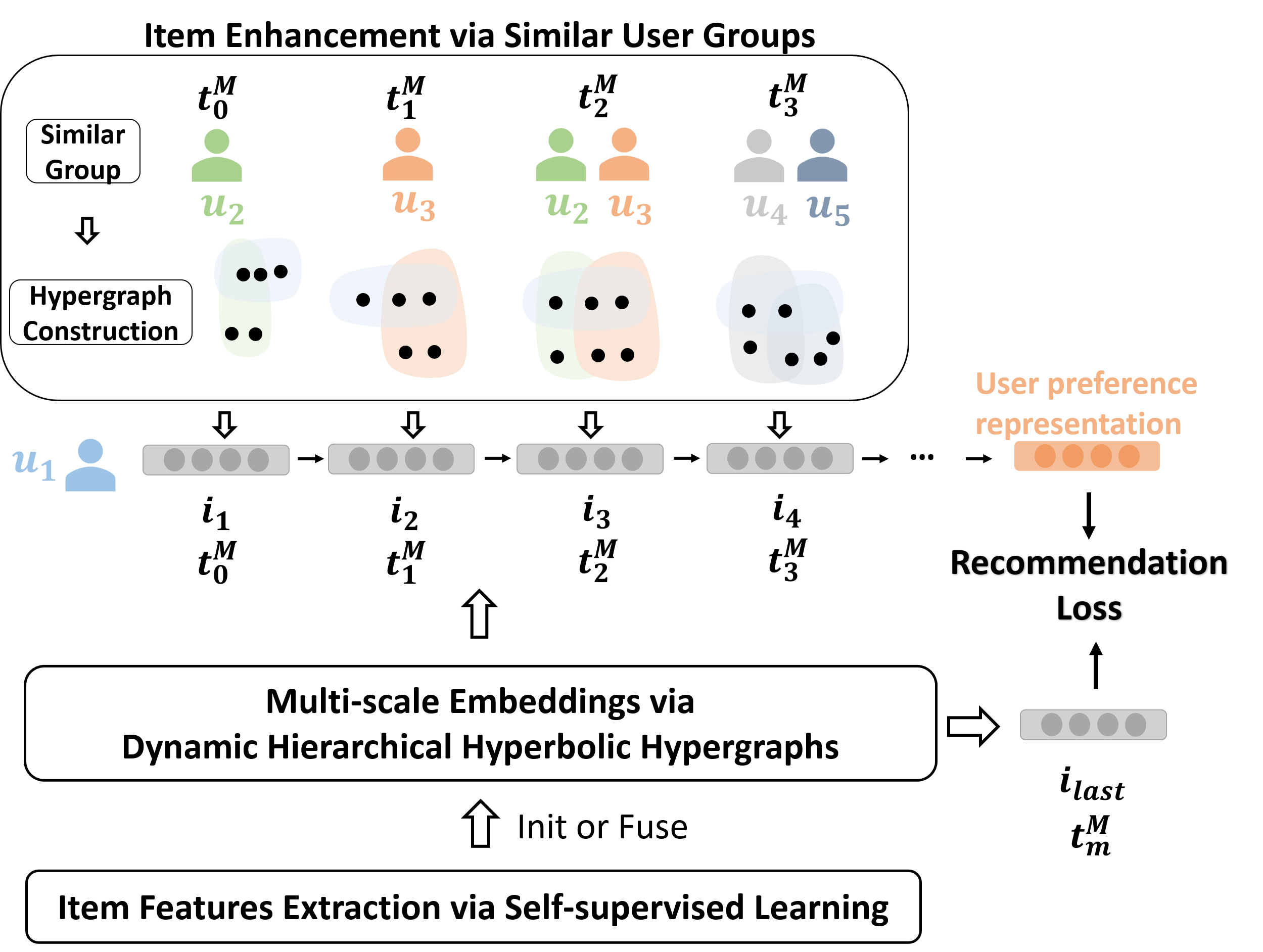}
	\vspace{-1em}
	\caption{Overall architecture $\mathbf{H^2 SeqRec}$ contains three modules, \textit{multi-scale embeddings via dynamic hierarchical hyperbolic hypergraphs}, \textit{item enhancement via similar user groups}, and \textit{learning user preference for sequential recommendation}. The pre-training module could be as initial embedding for $\mathbf{H^2 SeqRec}$ ($\mathbf{H^2 SeqRec}$-init); and it also could fuse into $\mathbf{H^2 SeqRec}$ ($\mathbf{H^2 SeqRec}$-fuse).
	}\vspace{-1em}
	\label{fig:architecture} \vspace{-1em}
\end{figure}

Figure \ref{fig:architecture} is an overview of our proposed sequential recommendation system.
Technically, we first generate yearly, quarterly, and monthly hypergraph snapshots as views from users' historical records. Here each hyperedge in the hypergraph refers to one user's historical items. 
With these hypergraphs, we design a novel hypergraph convolutional network in the hyperbolic space to learn multi-scale item embeddings. To enhance the item embedding, we further explore user's behaviour pattern via similar users' behaviours. In the end, we leverage the Transformer to learn users' preference and predict the next item via multi-layer perception (MLP). Moreover, before the $\mathrm{H^2 SeqRec}$, we design three self-supervised tasks, as shown in Figure \ref{fig:self-supervised-tasks}, to pre-train item representations. The pre-trained item representations could be as initial embeddings for $\mathrm{H^2 SeqRec}$, named $\mathrm{H^2 SeqRec}$-init; and they could also be fused into $\mathrm{H^2 SeqRec}$, named $\mathrm{H^2 SeqRec}$-fuse.

\vspace{-1em}
\subsection{Item Features Extraction via Self-supervised Learning}


\begin{figure*}
	\centering
	\begin{subfigure}[t]{0.32\textwidth}
		\centering
		\includegraphics[width=\linewidth]{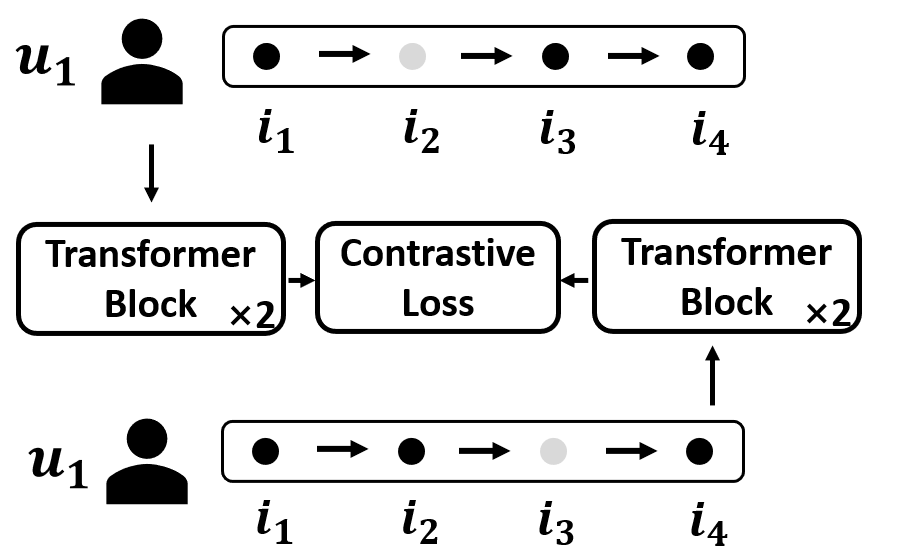} 
		\vspace{-1.5em}
		\caption{Mask random items}
		\vspace{-1em} 
		\label{fig:self-supervised-task1}
	\end{subfigure}
	\hfill
	\begin{subfigure}[t]{0.32\textwidth}
		\centering
		\includegraphics[width=\linewidth]{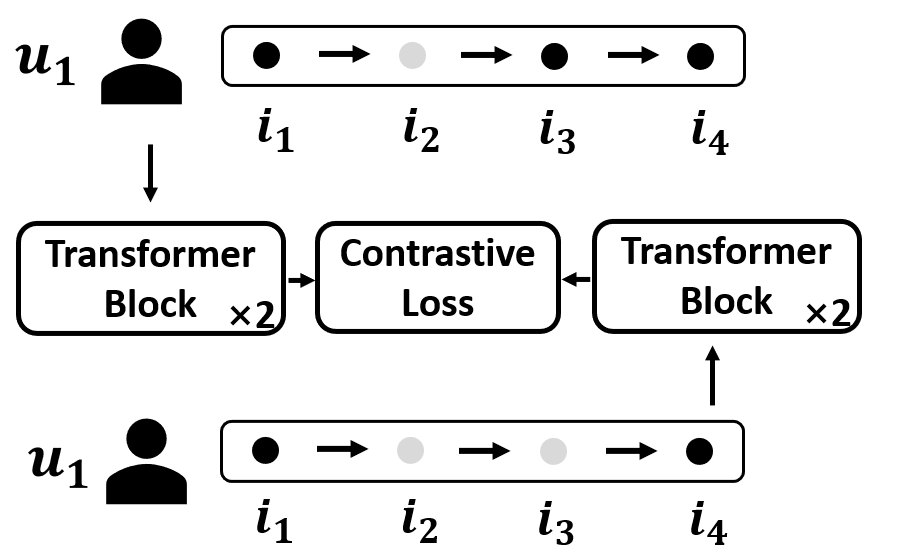} 
		\vspace{-1.5em}
		\caption{Mask subsequence items}
		\vspace{-1em}
		\label{fig:self-supervised-task2}
	\end{subfigure}
	\hfill
	\begin{subfigure}[t]{0.32\textwidth}
		\centering
		\includegraphics[width=\linewidth]{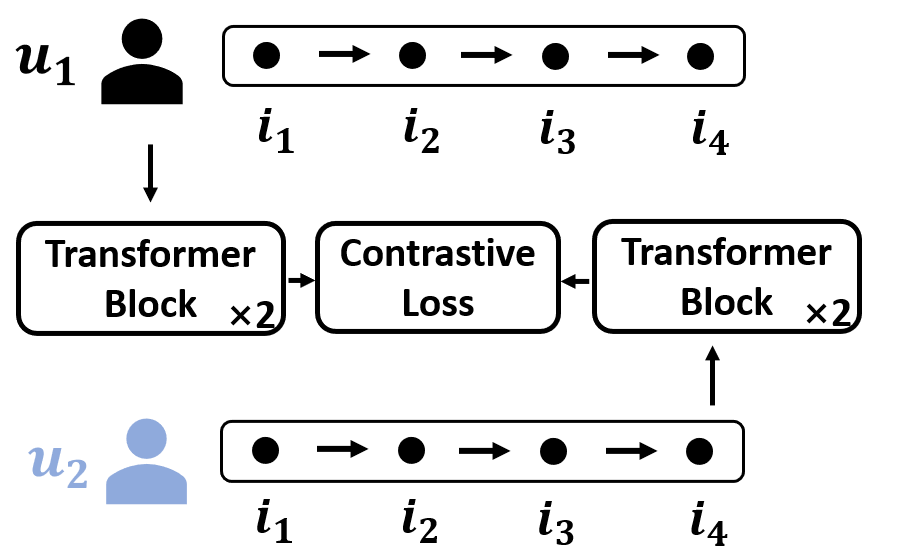} 
		\vspace{-1.5em}
		\caption{Hyperedge link prediction}
		\vspace{-1em}
		\label{fig:self-supervised-task3}
	\end{subfigure}

	\caption{Pre-training phase with three self-supervised tasks.}
	\vspace{-1em}
	\label{fig:self-supervised-tasks}
\end{figure*}


Before we start, we need the initial item features to support the downstream model. Inspired by the self-supervised learning framework \cite{zhou2020s3}, 
which can learn item representations without annotations, we propose three novel pre-training tasks (as shown in Figure \ref{fig:self-supervised-tasks}) to learn the pre-trained item representations:  

\begin{itemize}
    \item The first task is to predict whether a masked item sequence is derived from a user's historical records. The rationale is that in the real world, one item usually cannot change a user's long-term preference. Therefore, we randomly mask different items for a target user's historical sequence and then let these sequences' representations as close as possible. 
    \item The second task is to predict whether a sub-sequence is consistent with a user's short-term interest. The motivation is that the consecutive items taken in a short time usually contain a user's short-term interest, and these items might share similar functions. If we mask part of them, the user's short-term interest should be still observable from the rest items. Therefore, we first mask a sub-sequence with length 2 and then wish the masked sequence be close to the complete sequence. 
    \item The third task is to predict whether two users have the same behaviour patterns. Intuitively, users' behaviour patterns can be reflected in their historical records. If two users share many historical items, they are more likely to have the same preference, hobbies or even tastes. To this end, we use a hyperedge to connect one user's historical items and then let a pair of hyperedges be as close as possible if they share the overlapping items.  
\end{itemize}

Let $\mathbf{h_{s_u}}$ and $\mathbf{h_{s_u^{\prime}}}$ being embeddings of two sequences, we adopt contrastive loss \cite{xie2020contrastive} for each task:

\vspace{-1em}
\begin{equation}
\mathcal{L}\left(\!s_{u}, s_{u}^{\prime}, S_{u}^{-}\right)\!=\!-\log \!\frac{\exp \left(\operatorname{sim}\left(\mathbf{h_{s_u}}, \mathbf{h_{s_u^{\prime}}}\right) / \tau\right)}{\exp \left(\frac{\operatorname{sim}\left(\mathbf{h_{s_u}}, \mathbf{h_{s_u^{\prime}}}\right)}{\tau}\right)\!+\!\sum_{s_{u}^{-} \in S_{u}^{-}} \exp \left(\frac{\operatorname{sim}\left(\mathbf{h_{s_u}}, \mathbf{h_{s_u^{-}}}\right)}{\tau}\right)}
\end{equation}
here $S_{u}^{-}$ is the sampled negative sequences of the user $u$. $\mathbf{h_{s_u^{-}}}$ is the embedding of the negative sequence, $sim(\mathbf{h_{s_u}}, \mathbf{h_{s_u^{\prime}}})$ is defined as: $\frac{\mathbf{h_{s_u}^T} \cdot \mathbf{h_{s_u^{\prime}}}} {||\mathbf{h_{s_u}}|| \cdot ||\mathbf{h_{s_u^{\prime}}}||}$ where $\tau$ is the hyper-parameter. The sequence embeddings $\mathbf{h_{s_u}}$ and $\mathbf{h_{s_u^{\prime}}}$ are both calculated by Transformer network with two multi-head attention and feed forward blocks.
We combine the above three tasks together with optimized weights \{0.1,1,1\}.
\vspace{-1em}
\subsection{Multi-scale embeddings via dynamic hierarchical hyperbolic hypergraphs}
\label{subsec:hierarchical_item_emb}

\begin{figure}[t]
	\centering
	\includegraphics[width=0.45\textwidth]{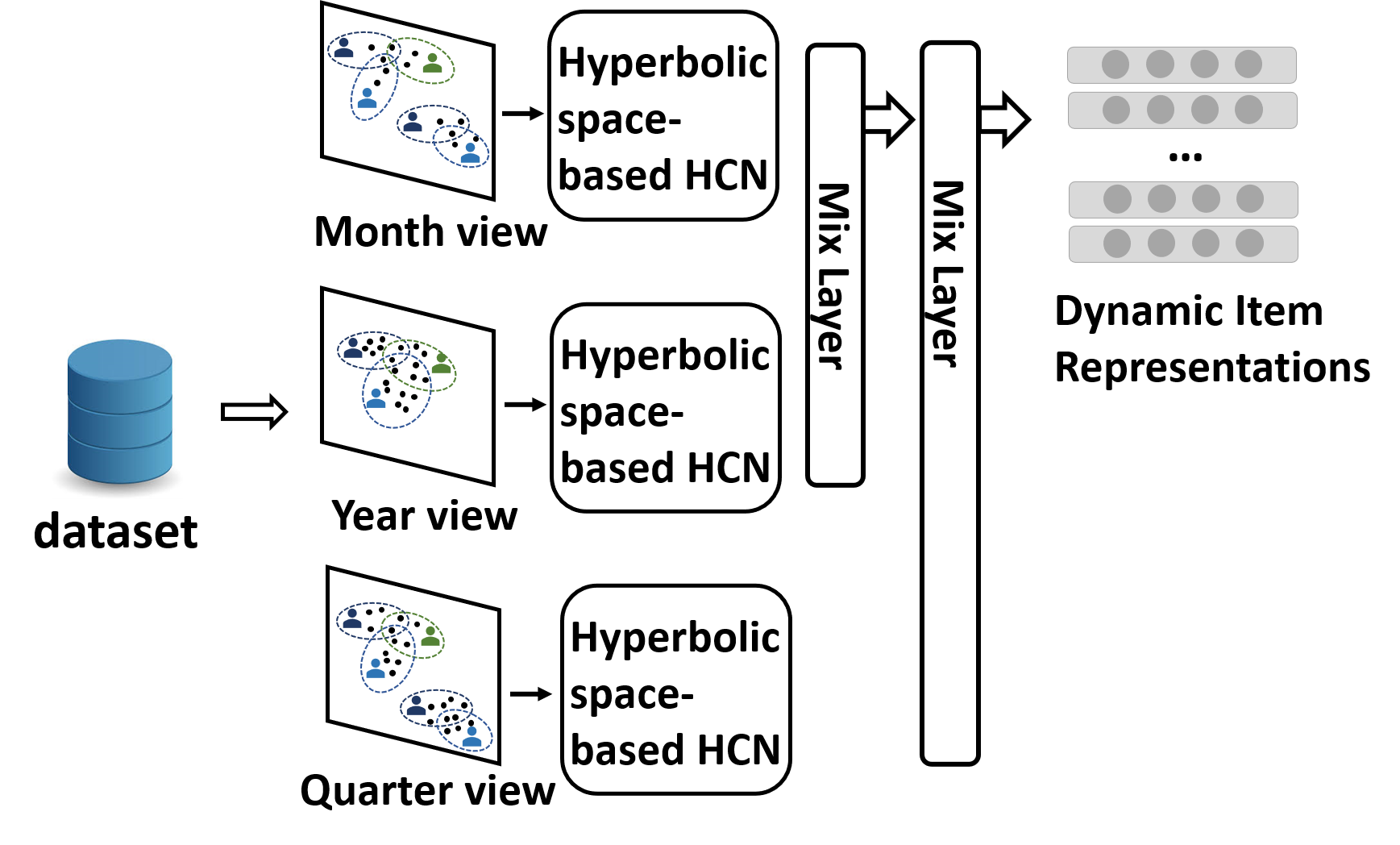}\vspace{-1em}
	\caption{Multi-scale Embeddings via Dynamic Hierarchical Hyperbolic Hypergraphs.} \vspace{-1.5em}
	\label{fig:hierarchical_with_hhcn} 
	
\end{figure}

With the above initial features, we then design a dynamic hypergraph neural network for learning the multi-scale item embeddings in hyperbolic space, as shown in Figure \ref{fig:hierarchical_with_hhcn}.
\vspace{-0.5em}
\subsubsection{Hierarchical Time-span Hypergraph Construction}
\label{sec:hierarchical}
To model the complex dependencies among the sequential items, we propose a hierarchical architecture to learn the monthly, quarterly and yearly relationship among sequential items. The motivation is that a user's behaviour patterns are not just suggested in anteroposterior items of the item sequences but also reflected via the seasonal and periodical variance. For example, users prefer to buy Christmas-related products before Christmas, and purchase T-shirts in summer and coat in winter. Inspired by \cite{wang2020next}, we utilize hypergraphs instead of traditional graphs to model different sub-sequences of each user and we also use multi-scale time-spans views (monthly, quarterly, and yearly) to learn the semantics of items. To tackle the sparsity and the long-tail distribution of these hypergraphs, we learn the graph representations in hyperbolic space because it is perfectly suitable for long-tail structures.

Specifically, when we deal with the monthly item representations, we first construct a monthly hypergraph where each hyperedge connects user's items within one month. Then we use hyperbolic space-based hypergraph graph neural network (introduced in \ref{subsec:hyperbolic_hypergraph_nn}) to learn the dynamic item embeddings in each month. Following this approach, we can also learn quarterly and yearly item representations.
\vspace{-1em}
\subsubsection{Hyperbolic Space-based Hypergraph Convolutional Network}
\label{subsec:hyperbolic_hypergraph_nn}
Compared with traditional graphs that mainly rely on pairwise user-item interactions, hypergraphs can model much higher relations in user-item interactions and promise to fuse item context to remit the sparsity problem. For example, if user $u_1$ bought flower and wedding dress and $u_2$ bought flower and a vase, we can use a hyperedge to connect the flower and the wedding dress, and use another hyperedge to connect the flower and the vase. In this way, the flower's semantic for $u_1$ is about the wedding but for $u_2$ is about decoration. However, if we use a traditional graph, the graph is hard to reveal different item semantics directly.

In addition, the user-item record usually follows the long-tail distribution and is sparse. The traditional Euclidean space usually cannot capture this structure leading to representation distortion, while hyperbolic space is beneficial to deal with the issue. \cite{chami2019hyperbolic}

Based on the above motivations, we propose a hypergraph neural network to model the items evolution in hyperbolic space. Specifically, we first transform the initial item features from Euclidean space to hyperbolic space $\mathbb{H}^{c}$, and then we feed the initial hyperbolic item embeddings to learn item embeddings. For the hyperbolic space, we set $\mathbf{o}:=\{\sqrt{c}, 0,0,...,0\}\in \mathbb{H}^{c}$ as the north pole in $\mathbb{H}^{c}$, where $-1/c$ is the negative curvature of hyperbolic model. As analyzed in \cite{chami2019hyperbolic}, the initial item features in hyperbolic space can be deduced from Euclidean space as follows:
\begin{equation}
\begin{split}
\mathbf{h}^{0, \mathbb{H}}&=\exp _{\mathbf{o}}^{c}\left(\left(0, \mathbf{h}^{0,\mathbb{E} }\right)\right)\\
&=\left(\sqrt{c} \cosh \left(\frac{\left\|\mathbf{h}^{0, \mathbb{E}}\right\|_{2}}{\sqrt{c}}\right), \sqrt{c} \sinh \left(\frac{\left\|\mathbf{h}^{0, \mathbb{E}}\right\|_{2}}{\sqrt{c}}\right) \frac{\mathbf{h}^{0, \mathbb{E}}}{\left\|\mathbf{h}^{0, \mathbb{E}}\right\|_{2}}\right)
\end{split}
\end{equation}
where $\mathbf{h}^{0, \mathbb{H}}$ and $\mathbf{h}^{0, \mathbb{E}}$ are the initial hyperbolic embedding and the initial Euclidean embedding, respectively. 

With the initial features in hyperbolic space, we further transform the features via a linear function in hyperbolic space:
\begin{equation}
\label{eq:hyperbolic_trans}
    \mathbf{x}_i^{L,\mathbb{H}} = (W^L \otimes^c \mathbf{h}_i^{L-1,\mathbb{H}}) \oplus^c \mathbf{b}^L
\end{equation}
where $\mathbf{x}_i^{L,\mathbb{H}}$ is the hyperbolic hidden embedding of item $i$ in the $L$-th layer after transformation, $W^L$ and $\mathbf{b}^L$ are the weight and bias, respectively. $\mathbf{h}_i^{L-1,\mathbb{H}}$ is item $i$'s hyperbolic embedding in the last layer. When $L=1$, $\mathbf{h}_i^{L-1,\mathbb{H}} = \mathbf{h}_i^{0, \mathbb{H}} $. Then, the item embeddings can be aggregated from the neighbouring nodes via the following convolutional operation in hyperbolic space:

\begin{equation}
\label{eq:hyperbolic_aggre}
    \mathbf{y}_i^{L,\mathbb{H}} = exp_{\mathbf{x}_i^{L,\mathbb{H}}}^{c}(\sum _{j\in \mathcal{N}(i)\cup j\in \mathcal{N}(i^+), where\ i^+ \in e_i^H} M_{ij} log_{\mathbf{x}_i^{L,\mathbb{H}}}^c (\mathbf{x}_j^{L,\mathbb{H}})
\vspace{-0.5em}
\end{equation}
where $\mathbf{y}_i^{L,\mathbb{H}}$ is the hyperbolic hidden embedding of item $i$ in the $L$-th layer after aggregation, $\mathcal{N}(.)$ denotes the neighbors so $\mathcal{N}(i)$ is node $i$'s neighbors and $\mathcal{N}(i^+)$ is node $i^+$'s neighbors where $i^+$ is on the same hyperedge as $i$. The node $j$'s hyperbolic embedding is transformed to Euclidean embedding via $log_\mathbf{o}^c(.)$, so the Euclidean-based sum and add operations are available. $exp^c(.)$ aims to transform the Euclidean-based embedding to hyperbolic embedding. According to \cite{chami2019hyperbolic}, choosing $\mathbf{x}_i^{L,\mathbb{H}}$ as the north pole is the best Euclidean approximation at this step. $M_{ij}$ is the 
projecting weight defined as follows: 
\begin{equation}
\label{eq:hyperbolic_aggre2}
    M_{ij} = \underset{j\in \mathcal{N}(i)\cup j\in \mathcal{N}(i^+), where\ i^+ \in e_i^H} {softmax} (MLP(log_{\mathbf{o}}^c(\mathbf{x}_i^{L,\mathbb{H}}) || log_{\mathbf{o}}^c(\mathbf{x}_j^{L,\mathbb{H}})))
\end{equation}

With the above node features' aggregation, we then use an activation function to generate the hidden embedding on Layer $L$:
\vspace{-0.5em}
\begin{equation}
    \mathbf{h}_i^{L,\mathbb{H}} = ReLU^{\otimes ^ c}(\mathbf{y}_i^{L,\mathbb{H}})
\end{equation}
where $\mathbf{h}_i^{L,\mathbb{H}}$ is the node $i$'s hyperbolic embedding at $L$-th layer, $ReLU^{\otimes ^ c}$ is the $ReLU$ operation in the hyperbolic space.
Through $log_\mathbf{o}^c(.)$, we transform the hyperbolic item embedding $\mathbf{h}_i^{2,\mathbb{H}}$ to Eucli-dean-space based embedding $\mathbf{h}_i^{\mathbb{E}}$, and feed into the next module.

\vspace{-0.5em}
\subsubsection{Mix Layer}
\label{sec:mix_layer}

With item embeddings for each month, quarter and year, we then design a mix layer to fuse them together. Specifically, to fuse monthly embeddings and yearly embeddings, we use a two-layer neural network to calculate their correlations as follows:
\begin{equation}
\label{eq:mix_layer_1}
	h_{MY_{t_m}} = ReLU(W_{MY}h_{Y_{t_y}}+\mathbf{b}_{MY})\odot h_{M_{t_m}}
\end{equation}
where $h_{MY_{t_m}}$ represents the mixture of monthly and yearly item embeddings at timestamp $t_m$. $h_{M_{t_m}}$ and $h_{Y_{t_y}}$ mean the monthly and yearly item embeddings respectively. $t_m$ and $t_y$ denote the monthly and yearly timestamp, respectively. $W_{MY}$ and $\mathbf{b}_{MY}$ are weight matrix and bias. $\odot$ means element-wise product. Similarly, the mixed embeddings with the month and the quarter can be defined as $h_{MQ_{t_m}}$, which is calculated by:
\begin{equation}
h^{t_m, Hie} = h_{MQ_{t_m}} = ReLU(W_{MQ}h_{Q_{t_q}}+\mathbf{b}_{MQ})\odot h_{MY_{t_m}}
\end{equation}
where $h^{t_m, Hie}$ is the dynamic item embeddings at timestamp $t_m$ learned by this hierarchical architecture.

\vspace{-1em}
\subsection{Item Enhancement via Similar User Groups}
\label{subsec:item_emb_enhance_via_user}
Most existing hypergraph-based sequential recommender systems \cite{wang2020next} ignore the hidden relationships among users, like social relationships, co-purchasing relationship, etc. However, these relationships are highly informative to capture the hidden users' behaviours. For example, a user's shopping behaviour may be affected by his friend circle, making the user tend to buy similar products with his friends.
In light of this, we utilize user groups to enhance the former learned dynamic item embeddings.

Specifically, we first find similar users for each user at each timestamp $t_m$. The similar users are those who take the same products at the same timestamp $t_m$. 
Using the current user's shopping records at $t_m$ and his similar users' records at $t_m$, we build an induced hypergraph to model his behaviour pattern in group-level. Each hyperedge connects each user’s items. Then we leverage our proposed hyperbolic space-based hypergraph convolutional network to learn item embeddings in group-level.

To enhance the item representations, we mix the above embeddings $\mathbf{h}_{iu}^{t_m, User}$ with previous embeddings $\mathbf{h}_{i}^{t_m, Hie}$ in Section \ref{sec:hierarchical}. The mix operation is defined as follows:
\begin{equation}
\label{eq:mix_user_hie}
	\mathbf{h}_{iu}^{t_m} = ReLU(W \mathbf{h}_{iu}^{t_m, User} + \mathbf{b}) \odot \mathbf{h}_{i}^{t_m, Hie}
\end{equation}
where $\mathbf{h}_{iu}^{t_m}$ is the item $i$'s embedding for user $u$ at timestamp $t_m$.

\vspace{-1em}
\subsection{Learning User Preference for Sequential Recommendation}
In this section, we leverage learned item embeddings to find users' preference and then present the sequential recommendation model.
\subsubsection{User Preference Learning}

For each user $u$, we use the Transformer \cite{vaswani2017attention} to model the dynamic item embeddings during the whole timestamps, mathematically,
\begin{equation}
	\mathbf{h}_{\mathcal{S}_u} = Transformer(\mathbf{h}_{iu}^{t_1}, \mathbf{h}_{iu}^{t_2}, ..., \mathbf{h}_{iu}^{t_m})  ,\ i\ \in\ \mathcal{S}_u
\end{equation}
where $\mathbf{h}_{\mathcal{S}_u}$ is the embedding of user $u$'s item sequence, which can describe $u$'s preference. The input of Transformer are the dynamic item embeddings learned in Equation \ref{eq:mix_user_hie}. In this way, the user's sequential items contain both the dynamic hierarchical information and the user's potential group information.

\subsubsection{The Complete Model}
With user preference representation $\mathbf{h}_{\mathcal{S}_u}$, we use a two-layer Multilayer Perceptron (MLP) to calculate the rating score of the next item, mathematically,
\begin{equation}
	r_{u,i}^{t_m} = MLP(\mathbf{h}_{\mathcal{S}_u}, \mathbf{h}_i^{t_m})
\end{equation}
where $r_{u,i}^{t_m}$ is the rating score of user $u$ and item $i$ at timestamp $t_m$, $\mathbf{h}_i^{t_m}$ is the embedding of item $i$ at timestamp $t_m$. We train the recommendation by Bayesian Pairwise Loss \cite{rendle2012bpr}, which aims to maximize the difference between the rating scores of the positive item $i$ and negative sample $j$:
\vspace{-0.5em}
\begin{equation}
loss = \sum _{(u,i,t,j)\in \mathcal{D}} -ln \sigma(r_{u, i}^{t_m}, r_{u, j}^{t_m}) + \alpha ||\delta||^2
\end{equation}
where $(u,i,t,j)\in \mathcal{D}$ denotes the positive pair (u,i,t) and the negative pair (u,j,t) from the training set $\mathcal{D}$. $\sigma$ is the sigmoid function, $\alpha$ is the weight of L2 regularization term $||\delta||^2$.


%% file: experiment.tex
\section{Experiment}

\begin{table}[]
	\begin{tabular}{cccc}
		\hline
		\textbf{Dataset} & \textbf{No. Users} & \textbf{No. Items} & \textbf{No. Interactions} \\ \hline
		AMT              & 5352              & 11672             & 59495                     \\ \hline
		Goodreads        & 16701             & 20823             & 1084781                   \\ \hline
	\end{tabular}
	\caption{Datasets Description.}
	\vspace{-3em}
	\label{dataset_table}
\end{table}

\subsection{Experiment Settings}

\subsubsection{Dataset}

We evaluate our method on two real-world datasets, named AMT and Goodreads. The statistics are shown in Table \ref{dataset_table}.  
\begin{itemize}
	\item \textbf{AMT.} It is the subset of the public Amazon dataset \cite{ni2019justifying}, which contains consumers' buying record and reviews in 29 categories. In our experiment, we choose three categories, Automotive, Musical Instruments and Toys and Games, to form the dataset and remove users bought less than 5 items. The time of AMT dataset is range from 2014 to 2018. The first two items in 2018 are considered as valid data and test data, respectively. Generally, the distribution of the items among users is long-tail distribution \cite{yin2020learning}. 
	\item \textbf{Goodreads.} It \cite{wan2018item, wan2019fine} is collected from goodreads website, which is an online book community website. We choose user-book (item) interactions between 2013 and 2015 as the dataset. Last two interactions are valid and test data, respectively. From Table \ref{dataset_table}, it is evident that Goodreads is much denser than AMT dataset.
\end{itemize}


\subsubsection{Baselines}
\begin{itemize}
	\item \textbf{GRU4Rec.} \cite{hidasi2018recurrent,hidasi2015session} It is a well-known sequential-based recommendation model that utilizes GRU to sequentially model users' interactions to achieve top-N recommendation. 
	

	\item \textbf{SASRec.} \cite{kang2018self} It is a self-attention based sequential approach for next item recommendation, which could capture each user's both long-term sequential relations through Point-Wise Feed-Forward Network and short-term interactions with items through an attention mechanism.
	
	\item \textbf{BERT4Rec.} \cite{sun2019bert4rec} This method employs deep bidirectional self-attention to sequentially model the user behaviours in two directions through Cloze task.
	
	\item \textbf{SRGNN.} \cite{wu2019session} The method utilizes graph neural networks to model the session sequences and obtain the complex item transitions for each session.
	
	\item \textbf{HGN.} \cite{DBLP:conf/kdd/MaKL19} It proposes a hierarchical gating network with the Bayesian Personalized Ranking in order to capture user's both long- and short-term interest.
	
	\item \textbf{HyperRec.} \cite{wang2020next}
	It uses sequential hypergraphs to model dynamic item embedding sequentially and fuses with static item embeddings as item representations. For each user, item representations are fed into Transformer network to obtain the next item recommendation.

\end{itemize}

\vspace{-1em}
\subsubsection{Evaluation}

Our proposed method focuses on recommending next item, and therefore we use Top $K$ Hit Ratio (HR@$K$) and Top $K$ Normalized Discounted Cumulative Gain (NDCG@$K$) as our evaluation metrics. We choose $K=\{1,5,10,20\}$ in the baseline comparison experiment. Our baseline HyperRec \cite{wang2020next} randomly selects 100 negative samples for each positive user-item pair. However, we think it is insufficient to reflect our model's effect accurately. Moreover, the baseline Bert4Rec and SRGNN's evaluation speed is too slow to test all the data.
Therefore, in our experiment, we randomly choose $\{100,500\}$ negative samples and rank $\{101,501\}$ items to calculate the HR@$K$ and NDCG@$K$ scores.

\subsubsection{Parameter Details}
We implement GRU4Rec, SASRec, \\BERT4Rec and SRGNN from the RecBole python package \cite{recbole}. For other methods, we use the public code provided by each paper. For all the methods, the feature dimension size is 100. For AMT dataset, all baselines' training batch size is set 512. We use the early stop function in RecBole whose condition is not updating NDCG@10 for 10 epochs. The hidden size is 100, and the dropout probability is 0.5, as the same as our proposed methods. Moreover, we choose BPR as the loss of GRU4Rec, SASRec, BERT4Rec and SRGNN to keep identical with our methods. For those baselines' code authors provided, we employ the given default parameters, but the learning rate is 0.001 for all baselines. For Goodreads dataset, some baselines' training time is so long with training batch size 512, so we change it to 4096 if the baseline does not run out of CUDA memory. Other parameters are the same as training AMT dataset. 

To evaluate our proposed model\footnote{https://github.com/Abigale001/h2seqrec}, we design three strategies to initialize it: 1) $\mathrm{H^2 SeqRec}$, one-hot item embeddings as input without pre-training tasks, 2) $\mathrm{H^2 SeqRec}$-init, pre-trained item embeddings as input and 3) $\mathrm{H^2 SeqRec}$-fuse, a combination of the first two, one-hot item embeddings as input, and pre-training item embeddings fused with dynamic item embeddings and group-level item embeddings.

For our proposed methods $\mathrm{H^2 SeqRec}$, $\mathrm{H^2 SeqRec}$-init and \\ $\mathrm{H^2 SeqRec}$-fuse, the parameter settings are the same. The layer of hyperbolic hypergraph convolutional network is 2, and we choose the hyperboloid model as hyperbolic geometry with the negative curvature -1. Moreover, since the Goodreads dataset is too dense, we sample $1/100$ users to find similar user groups. To simplify our proposed methods, we fuse the item embeddings learned via similar users' group with the dynamic item embeddings within each batch rather than for each user.
In Transformer, we set the number of heads and blocks, 2 and 1, respectively. The epoch of hyperbolic hypergraph convolutional network is 300. The maximum sequence is set to 50 for all the above methods. Other parameters are the same as the baselines'.

\subsection{Effectiveness Analysis on Baselines}

\begin{table*}[]
\resizebox{\textwidth}{60mm}{
\begin{tabular}{|c|c|c|c|c|c|c|c|c||c|c|c||c|}
\hline
Datasets                    & NEG & Metrics                  & GRU4Rec & SASRec & BERT4Rec & SRGNN  & HGN    & HyperRec & $\mathrm{H^2 SeqRec}$        & $\mathrm{H^2 SeqRec}$-init   & $\mathrm{H^2 SeqRec}$-fusc   & \textbf{Improvement} \\ \hline
\multirow{16}{*}{AMT}       & 100 & \multirow{2}{*}{NDCG@1}  & 0.0820  & 0.0878 & 0.0691   & 0.0889 & 0.0856 & 0.0871   & 0.1110          & 0.1153          & \textbf{0.1170} & 31.61\%     \\ \cline{2-2} \cline{4-13} 
                            & 500 &                          & 0.0247  & 0.0291 & 0.0349   & 0.0245 & 0.0284 & 0.0372   & 0.0413          & 0.0422          & \textbf{0.0435} & 16.94\%     \\ \cline{2-13} 
                            & 100 & \multirow{2}{*}{NDCG@5}  & 0.1725  & 0.1722 & 0.1530   & 0.1894 & 0.1672 & 0.1765   & 0.2107          & 0.2129          & \textbf{0.2135} & 12.72\%     \\ \cline{2-2} \cline{4-13} 
                            & 500 &                          & 0.0557  & 0.0600 & 0.0761   & 0.0603 & 0.0621 & 0.0753   & 0.0780          & \textbf{0.0834} & 0.0832          & 9.59\%      \\ \cline{2-13} 
                            & 100 & \multirow{2}{*}{NDCG@10} & 0.2187  & 0.2159 & 0.1978   & 0.2390 & 0.2019 & 0.2216   & 0.2568          & 0.2571          & \textbf{0.2589} & 8.33\%      \\ \cline{2-2} \cline{4-13} 
                            & 500 &                          & 0.0745  & 0.0766 & 0.0965   & 0.0830 & 0.0775 & 0.0968   & 0.0973          & 0.1041          & \textbf{0.1050} & 8.47\%      \\ \cline{2-13} 
                            & 100 & \multirow{2}{*}{NDCG@20} & 0.2689  & 0.2632 & 0.2488   & 0.2857 & 0.2333 & 0.2725   & 0.3013          & 0.3034          & \textbf{0.3054} & 6.90\%      \\ \cline{2-2} \cline{4-13} 
                            & 500 &                          & 0.0962  & 0.0950 & 0.1209   & 0.1058 & 0.0971 & 0.1159   & 0.1189          & 0.1263          & \textbf{0.1281} & 5.96\%      \\ \cline{2-13} 
                            & 100 & \multirow{2}{*}{HR@1}    & 0.0820  & 0.0878 & 0.0691   & 0.0889 & 0.0856 & 0.0871   & 0.1110          & 0.1153          & \textbf{0.1170} & 31.61\%     \\ \cline{2-2} \cline{4-13} 
                            & 500 &                          & 0.0247  & 0.0291 & 0.0349   & 0.0245 & 0.0284 & 0.0372   & 0.0413          & 0.0422          & \textbf{0.0435} & 16.94\%     \\ \cline{2-13} 
                            & 100 & \multirow{2}{*}{HR@5}    & 0.2601  & 0.2543 & 0.2367   & 0.2866 & 0.2448 & 0.2640   & 0.3087          & 0.3098          & \textbf{0.3111} & 8.55\%      \\ \cline{2-2} \cline{4-13} 
                            & 500 &                          & 0.0876  & 0.0914 & 0.1158   & 0.0962 & 0.0916 & 0.1129   & 0.1175          & \textbf{0.1265} & 0.1226          & 9.24\%      \\ \cline{2-13} 
                            & 100 & \multirow{2}{*}{HR@10}   & 0.4036  & 0.3903 & 0.3765   & 0.4415 & 0.3507 & 0.4045   & 0.4524          & 0.4469          & \textbf{0.4529} & 2.58\%      \\ \cline{2-2} \cline{4-13} 
                            & 500 &                          & 0.1463  & 0.1429 & 0.1794   & 0.1670 & 0.1362 & 0.1794   & 0.1775          & 0.1896          & \textbf{0.1913} & 6.63\%      \\ \cline{2-13} 
                            & 100 & \multirow{2}{*}{HR@20}   & 0.6033  & 0.5785 & 0.5796   & 0.6271 & 0.4735 & 0.6063   & 0.6291          & 0.6312          & \textbf{0.6385} & 1.82\%      \\ \cline{2-2} \cline{4-13} 
                            & 500 &                          & 0.2332  & 0.2167 & 0.2763   & 0.2577 & 0.2113 & 0.2556   & 0.2655          & 0.2805          & \textbf{0.2842} & 2.86\%      \\ \hline
\multirow{16}{*}{Goodreads} & 100 & \multirow{2}{*}{NDCG@1}  & 0.2496  & 0.2407 & 0.2639   & 0.2856 & 0.2325 & 0.2792   & \textbf{0.3154} & 0.3027          & 0.3070          & 10.43\%     \\ \cline{2-2} \cline{4-13} 
                            & 500 &                          & 0.0924  & 0.0894 & 0.1021   & 0.1298 & 0.1026 & 0.1212   & \textbf{0.1476} & 0.1357          & 0.1370          & 13.71\%     \\ \cline{2-13} 
                            & 100 & \multirow{2}{*}{NDCG@5}  & 0.4302  & 0.4124 & 0.4376   & 0.4588 & 0.3848 & 0.4576   & \textbf{0.4866} & 0.4772          & 0.4810          & 6.06\%      \\ \cline{2-2} \cline{4-13} 
                            & 500 &                          & 0.1911  & 0.1841 & 0.2032   & 0.2311 & 0.1899 & 0.2268   & \textbf{0.2584} & 0.2453          & 0.2456          & 11.81\%     \\ \cline{2-13} 
                            & 100 & \multirow{2}{*}{NDCG@10} & 0.4803  & 0.4637 & 0.4869   & 0.5057 & 0.4312 & 0.5044   & \textbf{0.5311} & 0.5229          & 0.5257          & 5.02\%      \\ \cline{2-2} \cline{4-13} 
                            & 500 &                          & 0.2320  & 0.2240 & 0.2431   & 0.2703 & 0.2243 & 0.2676   & \textbf{0.2971} & 0.2856          & 0.2854          & 9.91\%      \\ \cline{2-13} 
                            & 100 & \multirow{2}{*}{NDCG@20} & 0.5129  & 0.4989 & 0.5187   & 0.5349 & 0.4656 & 0.5359   & \textbf{0.5591} & 0.5517          & 0.5547          & 4.33\%      \\ \cline{2-2} \cline{4-13} 
                            & 500 &                          & 0.2700  & 0.2611 & 0.2785   & 0.3036 & 0.2559 & 0.3039   & \textbf{0.3311} & 0.3206          & 0.3206          & 8.95\%      \\ \cline{2-13} 
                            & 100 & \multirow{2}{*}{HR@1}    & 0.2496  & 0.2407 & 0.2639   & 0.2856 & 0.2325 & 0.2792   & \textbf{0.3154} & 0.3027          & 0.3070          & 10.43\%     \\ \cline{2-2} \cline{4-13} 
                            & 500 &                          & 0.0924  & 0.0894 & 0.1021   & 0.1298 & 0.1026 & 0.1212   & \textbf{0.1476} & 0.1357          & 0.1370          & 13.71\%     \\ \cline{2-13} 
                            & 100 & \multirow{2}{*}{HR@5}    & 0.5969  & 0.5723 & 0.5967   & 0.6160 & 0.5225 & 0.6207   & \textbf{0.6437} & 0.6391          & 0.6389          & 3.71\%      \\ \cline{2-2} \cline{4-13} 
                            & 500 &                          & 0.2875  & 0.2764 & 0.3008   & 0.3281 & 0.2674 & 0.3285   & \textbf{0.3632} & 0.3500          & 0.3505          & 10.56\%     \\ \cline{2-13} 
                            & 100 & \multirow{2}{*}{HR@10}   & 0.7513  & 0.7305 & 0.7486   & 0.7609 & 0.6653 & 0.7679   & \textbf{0.7807} & 0.7778          & 0.7779          & 1.67\%      \\ \cline{2-2} \cline{4-13} 
                            & 500 &                          & 0.4146  & 0.4005 & 0.4244   & 0.4498 & 0.3718 & 0.4553   & \textbf{0.4842} & 0.4765          & 0.4767          & 6.35\%      \\ \cline{2-13} 
                            & 100 & \multirow{2}{*}{HR@20}   & 0.8799  & 0.8686 & 0.8735   & 0.8758 & 0.8006 & 0.8895   & 0.8916          & \textbf{0.8921} & 0.8920          & 0.29\%      \\ \cline{2-2} \cline{4-13} 
                            & 500 &                          & 0.5652  & 0.5475 & 0.5649   & 0.5817 & 0.4936 & 0.5990   & \textbf{0.6195} & 0.6139          & 0.6152          & 3.42\%      \\ \hline
\end{tabular}}
\caption{Comparison with baselines. The last column is the improvement of the best proposed method than the best baseline.} \label{table:baselines}\vspace{-2em}
\end{table*}

\subsubsection{Overall Performance Analysis Between Different Baselines}
As shown in Table \ref{table:baselines}, we evaluate our proposed model with 6 state-of-the-art sequential recommendation baselines. Our proposed model $\mathrm{H^2 SeqRec}$, $\mathrm{H^2 SeqRec}$-init and $\mathrm{H^2 SeqRec}$-fuse could outperform all of them in both HR@$K$ and NDCG@$K$ in 100 and 500 negative sampling experiments.
For AMT dataset, our best model improves the best baseline 31.61\% and 16.94\% at NDCG@1 for negative sampling 100 and 500, respectively. In terms of Goodreads dataset, our best model outperforms the best baseline 10.43\% and 13.71\% at NDCG@1 for negative sampling 100 and 500, respectively. From the last column (Improvement), the advantages also could see in other evaluation metrics, especially in top 1 and 5 ranking. In recommendation task, it is significant to have better recommendation in top ranking, because proper recommendation in higher ranking means more effectiveness of the recommendation model.

\subsubsection{Effectiveness on Pre-train Features}

In Table \ref{table:baselines}, for AMT dataset, the proposed $\mathrm{H^2 SeqRec}$-fuse achieves the best performance and improves 5.41\% and	5.33\% than the NDCG@1 of $\mathrm{H^2 SeqRec}$ with 100 and 500 negative samples. While for the Goodreads dataset, $\mathrm{H^2 SeqRec}$ achieves the best performance on NDCG@1 and better than $\mathrm{H^2 SeqRec}$-fuse by 2.74\% and 7.74\% in negative sampling of 100 and 500, respectively. From the results, we could get the conclusion that when training sparser dataset, to do data augmentation as pre-training could help improve model. However, in terms of denser dataset, it is useless or even worse to do pre-training while our proposed model without pre-training could still achieve quite well performance. 
Therefore, unlike traditional recommendation models which blindly pursue pre-training to improve the model, our experiment shows that sometimes pre-training cannot really help improvement.

Moreover, from the columns $\mathrm{H^2 SeqRec}$-init and $\mathrm{H^2 SeqRec}$-fuse of Table \ref{table:baselines}, we find that the results are of small differences in two datasets. It means that no matter how to use the pre-training embedding in the model, it is almost the same. Therefore, the conclusion is that pre-training does play a part in the model.

\subsection{Improvement of Different Component for Overall Performance}


\begin{table}[]
\resizebox{0.45\textwidth}{12mm}{
\begin{tabular}{|c|c|c|c|c||c|}
\hline
                                            & $K=1$          & $K=5$           & $K=10$          & $K=20$ & Diff. \\ \hline
$\mathrm{H^2 SeqRec}$                       & \textbf{0.111} & \textbf{0.2107} & \textbf{0.2568} & \textbf{0.3013}                                            & -       \\ \hline
$\mathrm{H^2 SeqRec}$ $\neg$ U              & 0.1102         & 0.2083          & 0.2519          & 0.3011                                                     & 1.91\%  \\ \hline
$\mathrm{H^2 SeqRec}$ $\neg$ Hie            & 0.1084         & 0.2028          & 0.2466          & 0.2927                                                     & 3.96\%  \\ \hline
$\mathrm{H^2 SeqRec}$ $\neg$ HB                              & 0.1099         & 0.2048          & 0.2486          & 0.2961                                                     & 3.19\%  \\ \hline
$\mathrm{H^2 SeqRec}$($\mathrm{Hie^{MQY}}$) & 0.1095         & 0.2057          & 0.2509          & 0.2977                                                     & 2.30\%  \\ \hline
\end{tabular}}
\caption{Impact of Different Modules ($\mathbf{NDCG@K}$). $\neg$ is removing the following module and keeping the others. The last column means the decrease rate of each row compared with $\mathbf{H^2 SeqRec}$ on $\mathbf{NDCG@10}$.}
\label{table:ablation_modules}
\vspace{-3em}
\end{table}

We perform ablation test on AMT dataset to study different modules' effects. The evaluation metric is NDCG@\{1,5,10,20\}, and the number of negative test sampling is 100. 
In Table \ref{table:ablation_modules}, the $\neg$ represents removing the following module and remaining other modules. For example, $\mathrm{H^2 SeqRec}$ $\neg$ U means the proposed model without Item Enhancement via Similar User Groups module in Section \ref{subsec:item_emb_enhance_via_user}. In the next row, Hie means the hierarchical module, so removing the hierarchical module represents deleting the quarterly and yearly dynamic item embeddings and only feeding monthly dynamic ones into the recommendation. HB means hyperbolic space, and therefore that row means $\mathrm{H^2 SeqRec}$ abandons hyperbolic space and train in Euclidean space. $\mathrm{H^2 SeqRec}$($\mathrm{Hie^{MQY}}$) changes the order of fusion time-span in hierarchical architecture, that is, it fuses quarterly dynamic item embeddings firstly into monthly dynamic item embeddings, and then yearly item embeddings fuse into them. 

As shown in the Table \ref{table:ablation_modules}, the proposed model $\mathrm{H^2 SeqRec}$ achieves the best performance. The Item Enhancement via Similar User Groups module plays the least important role in the whole model, because of the sparsity of AMT dataset. Moreover, using the hierarchical module to learn dynamic item embeddings and hyperbolic space-based item embeddings learning both could improve the model. However, if we change the hierarchical time-span fusion order, the results will be a little bit worse. That may be because if yearly dynamic item embeddings fuse in the last step in hierarchical architecture, the final fused item embeddings would contain more yearly information and less quarterly information. However, yearly information may be too large as a time interval, and the quarterly one may be apposite. Therefore, that is why we choose to fuse yearly information first and then quarterly, and the results also prove our rationality. 

\vspace{-1em}
\subsection{Analysis of Different Pre-training Tasks Influence on Performance}
In Table \ref{table:ablation_pretrain}, we evaluate different pre-training tasks' influence on the performance of the AMT dataset with 100 negative samples. Minus after $\mathrm{H^2 SeqRec}$-fuse denotes only considering the following pre-training tasks. $M$ means the masking random items task, $S$ represents the masking subsequence task, and $H$ is the hyperedge link prediction task. 

$\mathrm{H^2 SeqRec}$-fuse-SH and $\mathrm{H^2 SeqRec}$-fuse-MH obtain better performance, compared with $\mathrm{H^2 SeqRec}$-fuse-MS, which proves the effectiveness of our proposed hyperedge prediction task. We can also get this conclusion by the last row, $\mathrm{H^2 SeqRec}$-fuse-H shows the highest performance compared with $\mathrm{H^2 SeqRec}$-fuse-M and $\mathrm{H^2 SeqRec}$-fuse-S. Therefore, this experiment shows our proposed hyperedge link prediction task's effectiveness, which is more helpful than the existing pre-training methods merely model each user's buying history in a chronological way.



\begin{table}[h!]
\begin{tabular}{|l|c|c|c|c|}
\hline
                 & $K=1$             & $K=5$             & $K=10$            & $K=20$            \\ \hline
$\mathrm{H^2 SeqRec}$-fuse    & \textbf{0.1170} & \textbf{0.2135} & \textbf{0.2589} & \textbf{0.3054} \\ \hline
$\mathrm{H^2 SeqRec}$-fuse-SH & 0.1149          & 0.2103          & 0.2568          & 0.3014          \\ \hline
$\mathrm{H^2 SeqRec}$-fuse-MH & 0.1158			
          & 0.2102          & 0.2544         & 0.3036          \\ \hline
$\mathrm{H^2 SeqRec}$-fuse-MS & 0.1138          & 0.2130          & 0.2584          & 0.3043          \\ \hline
$\mathrm{H^2 SeqRec}$-fuse-M  & 0.1123          & 0.2111          & 0.2541          & 0.3015          \\ \hline
$\mathrm{H^2 SeqRec}$-fuse-S  & 0.1117          & 0.2080          & 0.2533          & 0.3006          \\ \hline
$\mathrm{H^2 SeqRec}$-fuse-H  & 0.1140          & 0.2129          & 0.2572          & 0.3020          \\ \hline
\end{tabular}
\caption{Impact of different pre-training tasks ($\mathbf{NDCG@K}$). $-$ denotes only considering the following pre-training tasks.}
\label{table:ablation_pretrain}
\vspace{-2em}
\end{table}

\subsection{Parameters Sensitivity Study}
\begin{figure}[h]
	\centering
	
	\begin{subfigure}[t]{0.25\textwidth}
		\centering
		\hspace{-2em}
		\includegraphics[scale=0.25]{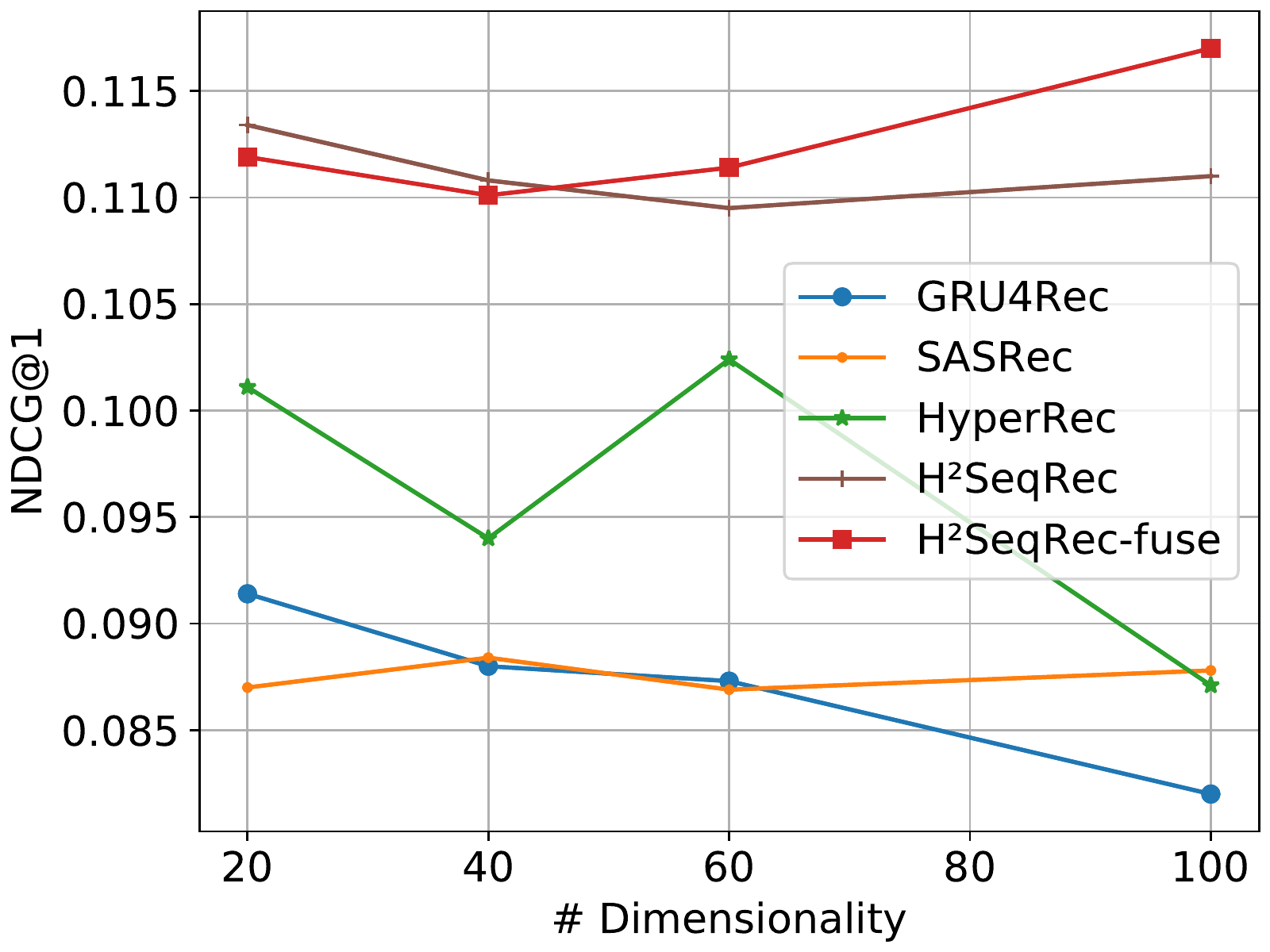} 
		\caption{NDCG@1 w.r.t. \# Dimensions} \label{fig:param_study_v1}
	\end{subfigure}
	~
	\begin{subfigure}[t]{0.25\textwidth}
		\centering
		\hspace{-2em}
		\includegraphics[scale=0.25]{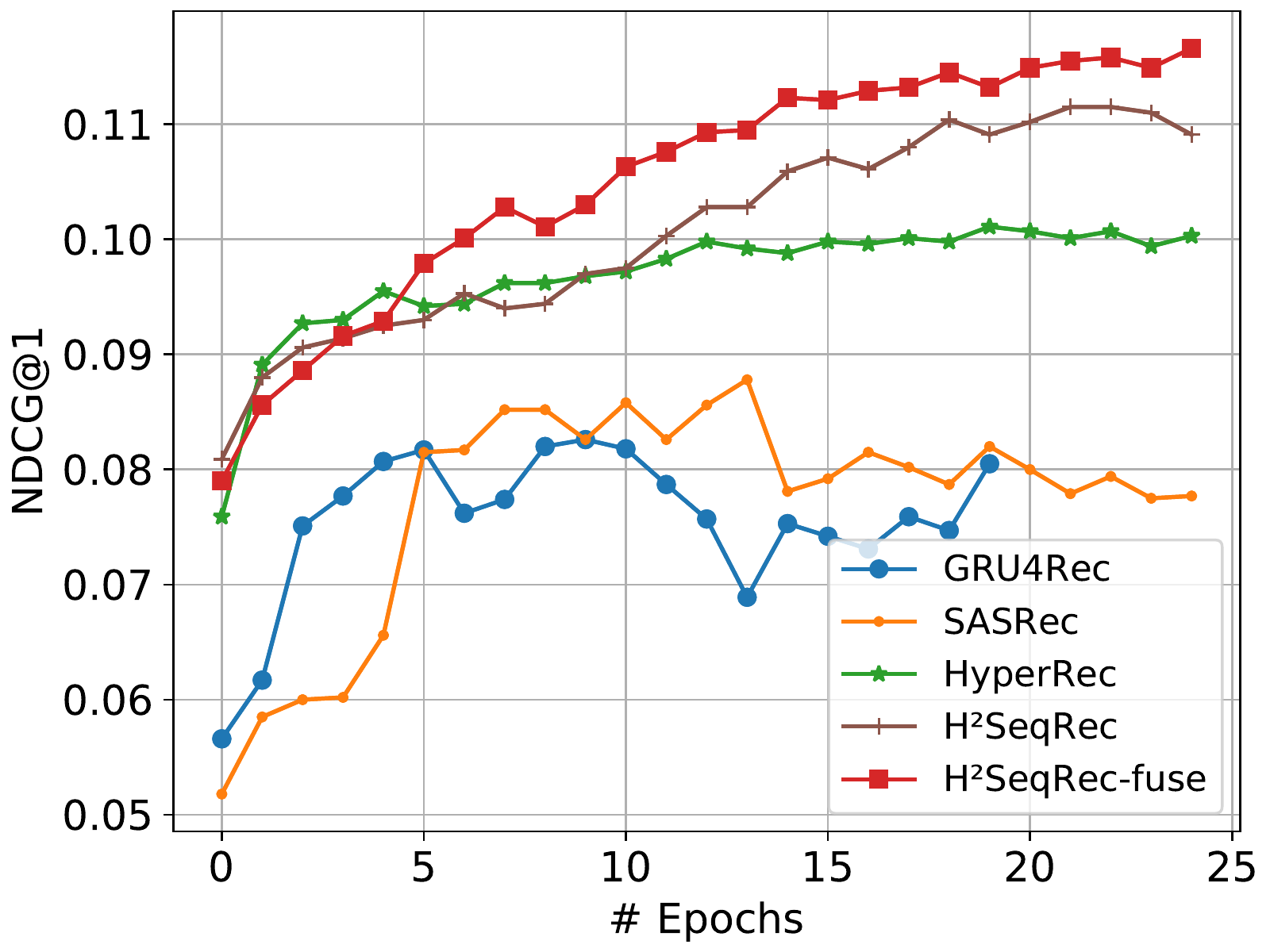} 
		\caption{NDCG@1 w.r.t. \# Epochs} \label{fig:param_study_v2}
		
	\end{subfigure}
	\vspace{-1em}
    \caption{Parameters Sensitivity.} \label{fig:parameter_sen}
    \vspace{-1em}
\end{figure}

In the Figure \ref{fig:parameter_sen}, we analysis different dimensions of items influence on the proposed model and the baselines, we randomly choose three baselines to compare with our proposed models. The negative test sampling is 100, and the evaluation metric is NDCG@1. 

As shown in Figure \ref{fig:param_study_v1}, our proposed models $\mathrm{H^2 SeqRec}$ and $\mathrm{H^2 SeqRec}$-fuse could always outperform GRU4Rec, SASRec and HyperRec when the dimension is \{20, 40, 60, 100\}. Moreover, when the number of dimensions is larger in the figure, the pre-training phase is more useful to the recommendation performance. It is because the pre-training data augmentation can learn more information about the sparse dataset when the number of dimensions is larger. 

In Figure \ref{fig:param_study_v2}, we compare the NDCG@1 on models training. GRU4Rec is early stopped because of not updating NDCG@1 for 10 epochs, so the line is not complete. In general, our proposed two methods always achieve the best performance, especially the $\mathrm{H^2 SeqRec}$-fuse, whose NDCG@1 is still updating after 25 epochs.

\vspace{-1em}
\subsection{Complexity Analysis}

Our model contains four parts, item features extraction via self-supervised learning tasks, learning multi-scale embeddings via dynamic hierarchical hyperbolic hypergraphs, item embeddings enhancement via similar users groups and learning user preference for sequential recommendation. 
Therefore, we will analyse the time complexity of four separate parts, respectively. 

For the self-supervised learning phase, we have three tasks. The first task, for each user, we should randomly select two different items to mask, so the time complexity is $O(U\times I_U\times (I_U-1))$, where $U$ is the number of users, and $I_U$ is each user's related items. Since $I_U \ll U$, the time complexity could approximately be $O(U)$. The second task is that we randomly mask a subsequence for each user, and the length is 2, so the time complexity is about $O(U\times I_U) \approx O(U)$. For the last task, we should find each user's neighbours and two-hop neighbours, and then build the hyperedge connecting each user's items within a timestamp. Therefore, if there are $T_M$ timestamps, the time complexity is $O(U\times T_M \times \mathcal{N}_U \times \mathcal{N}_{\mathcal{N}_U})$, where $\mathcal{N}_U$ is the average numbers of each user's neighbours and $\mathcal{N}_{\mathcal{N}_U}$ is the average number of each user's two-hop neighbours. Since $T_M \ll U, \mathcal{N}_U \ll U, \mathcal{N}_{\mathcal{N}_U} \ll U$, the complexity is approximately $O(U)$. The pre-training tasks' time complexity can be added together, and could approximately ignore the smaller magnitude terms, so the overall time complexity of pre-training is $O(U)$. In practice, it is unnecessary to do the pre-training phase every time. The learned pre-training features could be learned offline and stored. Even if new data comes, it can implement incremental training. Therefore, the time complexity is acceptable.

For the multi-scale dynamic hierarchical hyperbolic hypergraphs learning, we have three time-spans views including month, quarter, and year. For monthly hyperbolic hypergraph neural network learning, we consider each user's items within a month connected by a hyperedge. We perform hyperbolic-based hypergraph convolutional neural network on it, so the time complexity is $O(T_M\times I\times \mathcal{N}_I \times L \times D^2)$, where $\mathcal{N}_I$ is the average number of neighbours for each hyperedge connected to the item, because of the aggregation function in the model. $L$ is the number of layers, and $D$ is the embedding's dimension. The time complexity is the same as the quarterly and yearly model, but the number of quarter $T_Q$ and year $T_Y$ is less than $T_M$. To sum up, this module's time complexity is $O((T_M+T_Q+T_Y)\times I\times \mathcal{N}_I \times L \times D^2) \approx O(I)$, because $T_M,T_Q,T_Y, L$ are countable, the average number of item's neighbours is $\mathcal{N}_I \ll I$ and the dimension is set as 100 in our model. What's more, this module can also learn offline due to our step-by-step learning style. In practice, this will significantly save the time of the recommendation. The learning strategy is the same as the pre-training phase. If there are new data, this module can also do incremental training. Therefore, the time complexity of this module seems good.

For the item embeddings enhancement via similar users groups module, the training method and the hypergraph construction strategy is the same as the last module. However, this module focuses on the relationship between users and will construct multiple hypergraphs for each user. Therefore, the time complexity is $O(T_M\times U\times \mathcal{N}_{I_U}\times \mathcal{N}_{\mathcal{N}_{I_U}} \times L \times D^2)$, where $I_U$ means each user's related items, $\mathcal{N}_{I_U}$ is the average number of neighbours of each user's related items and $\mathcal{N}_{\mathcal{N}_{I_U}}$ is our defined similar user's related items in section \ref{subsec:item_emb_enhance_via_user}. Similarly, the approximate time complexity is $O(U)$ because $\mathcal{N}_{I_U} \ll U$, $\mathcal{N}_{\mathcal{N}_{I_U}} \ll U$. This module can also train offline in advance, thanks to our step-by-step training strategy. In practice, if there are new users fed into the model, it is convenient to train new users' item embeddings. If some existing users are fed into the model again, it is convenient to train incrementally based on the previous version of item embeddings.

In the last module, learning user preference for sequential recommendation module, we use the Transformer to obtain each user's preference embedding. For each user, we feed user's sequence into the Transformer (self-attention layer), and therefore the time complexity is $O(U\times I_{U}^2 \times D) \approx O(U)$ because of $I_{U}\ll U$ as claimed before. 
To train the recommendation, we sample 1 negative training item to maximize the difference between the positive user-item pair score and the negative one. The recommendation part's time complexity is $O(I_{U} \times (NEG+1))$. Since the user preference embedding learning part and the recommendation part are learned end to end, the whole time complexity of the last module is $O(U\times I_{U}^2 \times D \times I_{U} \times (NEG+1)) \approx O(U)$ because of $I_{U} \ll U$ and $NEG \ll U$.

To sum up, the final time complexity of the proposed model is the maximum time complexity of the above modules, which is $max(O(U), O(I))$, so it is acceptable.

%% file: related-work.tex
\vspace{-0.5em}
\section{related work}
\label{sec_related_work}

\subsection{Non-graph sequential modelling for neural recommendation}
Early neural recommendations building on typical deep neural networks mostly use recurrent neural networks (RNN) or convolutional neural networks (CNN) to capture the temporal patterns from users' historical records. Specifically, Quadrana et al. \cite{quadrana2017personalizing} design a multi-level RNN structure for learning temporal patterns in a sequence. Although RNN-based methods have their advantages in sequential learning, user-item interactions usually contain noisy information. Only learning these relations is far from achieving more reliable recommendation system. Unlike RNN-based method, Yuan et al. \cite{yuan2019simple} use CNN to learn from user-item sequences because CNN-based methods can not only capture long-term dependencies but also have the character of translation invariance, making the model more stable for various sequential orders. Later, attention is introduced in recommendation models \cite{sun2019bert4rec, ma2019hierarchical, song2019session, wu2019dual}. In particular, Kang et al. \cite{kang2018self} treat the user-item sequence as a sentence and model the temporal relations via a transformer model \cite{vaswani2017attention}. The transformer takes a self-attention unit to translate sequences as entity embeddings and position embeddings, which can be used to the downstream recommendation system.

\subsection{Graph-based sequential modelling for neural recommendation}
In the real world, however, users and items contain more non-linear relations with different topological structures. To this end, graph-based sequential modelling approaches become more and more popular to model the recommendation data. By modelling graph structure data, graph neural networks (GNN) are introduced recently. As a conceptional extension of linear CNNs, GNN-based method \cite{wu2019session, xu2019graph, huang2020uber} take the directed graph as input, and capture the interdependence directly on the graph. For example, Wu et al. \cite{wu2019session} treat each user or item as a node in the graph, and transform the user-item sequence as a path. With GNN-based model, they can learn both users and items embeddings over the whole graph. 

Later, more advanced technologies have been introduced such as, self-supervised learning \cite{jin2020self, xie2020contrastive, xin2020self, wu2020self, yao2020self}, group awareness \cite{wang2020group, huang2020efficient, ji2018gist} and so on. In particular, Hwang et al. \cite{hwang2020self} propose self-supervised auxiliary learning tasks to predict meta-paths to capture rich-information of a heterogeneous graph, and thereby improve the primary task to do link prediction.

\vspace{-1em}
\subsection{Hypergraph-based sequential modelling for neural recommendation}
Recently, instead of traditional graph, constructing hypergraphs to learn the data structure to do recommendation become a popular approach. 
Yu et al. \cite{yu2021self} design a multi-channel hypergraph convolutional network to enhance social recommendation by exploiting high-order user relations, which shows great improvement. However, it ignores the sequential information for users. 
Xia et al. \cite{xia2020self} model session data as a hypergraph and propose a dual channel hypergraph convolutional network for session-based recommendation. 
Wang et al. \cite{wang2021session} also construct hypergraphs for each session to model the item correlations, and they also introduce a hypergraph attention layer to flexibly aggregate correlated items in the session and infer next interesting item.
Although the above two methods both consider some temporal information, it is within a specific session, not a whole sequence. 
Wang et al. \cite{wang2020next} propose a novel next-item recommendation framework empowered by sequential hypergraphs to incorporate the short-term item correlations while modeling the dynamics over time and across users. The advantage is to consider sequential information, but all the above three hypergraph-based methods ignore the severe sparsity problem of hypergraphs.